\documentclass{cjaa}                    
\usepackage{graphicx}                   
\input{epsf.sty}                        
\input{psfig.sty}                       
\newcommand{\Msolar}{\mbox{\,$\rm M_{\odot}$}}        

\begin{document}
\title{Difference of $\alpha$-disks between Seyfert 1 galaxies and Quasars}
\author
{Wei-Hao Bian \inst{1,2}, Hong Dong \inst{3} and Yong-Heng Zhao
\inst{1}}
\institute{National Astronomical Observatories, Chinese
Academy of
 Sciences, Beijing 100012, China; whbian@lamost.bao.ac.cn\\
\and Department of Physics, Nanjing Normal University, Nanjing
210097, China \\
\and Department of Physics, Shangqiu Teachers College, Shangqiu
476000, China}

\date{Received ...; accepted ...}
\titlerunning{}
\authorrunning{W. Bian, H. Dong \& Y. Zhao}

\abstract{In a previous paper (Bian \& Zhao 2002), it was
suggested that contamination to the luminosity of galactic nucleus
from the host galaxies play an important role in determining
parameters of the standard $\alpha$ disk for AGNs. Using the
nuclear absolute B band magnitude instead of the total absolute B
band magnitude, the central black hole masses, the accretion rates
and the disk inclinations to the line of sight for 20 Seyfert 1
galaxies and 17 Palomar-Green (PG) quasars were recalculated. It
was found that small value of $\alpha$ is needed in the accretion
disk for Seyfert 1 galaxies compared with PG quasars. The
difference of $\alpha$ maybe lead to the different properties
between Seyfert 1 galaxies and Quasars. Furthermore, we found most
of the objects in this sample are not accreting at super-Eddington
rates when we adopted the nuclear optical luminosity in our
calculation. 
          }
\maketitle
\section{Introduction}

The standard paradigm of AGNs is an accretion disk surrounding a
central super-massive black hole. The wide emission lines are
coming from the material located in the regions outside the
accretion disk called broad line regions (BLRs). With the
reverberation mapping method, the sizes of BLRs from the center of
AGNs are obtained in 37 nearby AGNs (Ho 1998; Wandel et al. 1999;
Kaspi et al. 2000). In our one previous paper (Bian \& Zhao 2002),
we assume that the gravitational instability of standard thin
disks leads to the formation of BLRs, the B band luminosity comes
from standard thin disk and the motion of BLRs is virial. Using
the accretion disk theory, we determined some parameters of the
accretion disk in AGNs, such as the central black hole mass,
accretion rates, disk inclinations to the line of sight, $\alpha$
parameter. It was pointed out that the host contribution to the
optical luminosity of AGNs had a strong effect on the
determination of these parameters of the accretion disk.

Ho \& Peng (2001) used the nuclear luminosity in optical and radio
bands to discuss the radio loudness of Seyfert nuclei and found
that the majority of type 1 Seyfert nuclei is in the category of
radio-loud AGNs. They also found that a strong correlation between
the nuclear optical magnitude ($M^{nuc}_{B}$) and H$\beta$
luminosity ($L_{H\beta}$) (See their figure 6). Ho (2002) also
used this correlation to estimate the nuclear optical luminosity
to investigate the relations between radio luminosity, radio
loudness, and the black hole mass.

Here we adopted the nuclear absolute B band magnitude
($M^{nuc}_{B}$) from the literature (Schmidt \& Green 1983; Ho \&
Peng 2001; Ho 2002) and did the recalculation to obtain the
accretion disk parameters according the method introduced in the
paper Bian \& Zhao (2002). This method used in the paper of Bian
\& Zhao (2002) was briefly introduced in section 2. The
recalculated results are presented in section 3. Section 4 is our
conclusion. All cosmological calculations in this paper assume
$H_{0}=75 km s^ {-1}, \Omega =1.0, \Lambda=0$.

\section{The method}
\subsection{Formulae}

Here we briefly introduced the method used in Bian \& Zhao (2002).
First, using the standard thin disk theory, the B land luminosity
($L^{B}$) can be derived from the black hole mass ($M$), the
accretion rate ($\dot{M}$), and the inclination ($i$) by
\begin{equation}
L^{B}_{9}=13.8\dot{M}_{26}^{2/3}M_{8}^{2/3}cosi.
\end{equation}
where $M_{8}=M/(10^{8}\Msolar)$, $L^{B}_{9}=L^{B}/(10^{9}L\sun)$,
and $\dot{M}_{26}= \dot{M}/(10 ^{26} \rm{g~s^{-1}})$.

Second, from accretion disk theory, we can obtain the radius of
the gravitational instability ($R_{ins}$) (Bian \& Zhao 2002) by
\begin{equation}
R_{14}=880\alpha^{28/45}Q^{-8/9}\dot{M}_{26}^{-22/45}M_{8}^{1/3}.
\end{equation}
where $R_{14}=R_{ins}/(10^{14}cm)$, $\alpha$ is the parameter
describing the viscosity for the standard $\alpha$-disks. $Q$ is
the criterion of the gravitational instability. Assuming that the
gravitational instability of the accretion disk leads to BLRs, we
assume that the BLRs size ($R_{BLR}$) is equal to instability
radius ($R_{ins}$).

Third, assuming BLRs motion is virial, the FWHM of H$\beta$
($V_{FWHM}$) is given by
\begin{equation}
V_{3}=3.89Q^{4/9}\alpha^{-14/45}\dot{M}_{26}^{11/45}M_{8}^{1/3}sini.
\end{equation}
where $V_{3}=V_{FWHM}/(1000 \rm{km~s^{-1}})$.

Using Eq. (1-3), we can calculate the central black hole masses
($M$) , the accretion rates ($\dot{M}$), and the inclinations
($i$) knowing absolute B band luminosity ($L^{B}$), sizes of the
BLRs ($R_{BLR}$) and FWHMs of H$\beta$ ($V_{FWHM}$).

\subsection{Date}
Our sample consists of all AGNs with available BLRs
sizes from the reverberation mapping method: 20 Seyfert 1 galaxies
(Wandel et al. 1999, Ho 1998), 17 PG quasars (Kaspi et al 2000).
For this sample, $M^{nuc}_{B}$ is available in the literature for
all the quasars (Schmidt \& Green 1983) and for a number of the
Seyfert 1 galaxies (Ho \& Peng 2001). For the other Seyfert 1
galaxies, the $M^{nuc}_{B}$ is obtained from the
$M^{nuc}_{B}-L_{H\beta}$ correlation (Ho 2002). We used
$M^{nuc}_{B}$ instead of the absolute B band magnitude from the
Veron-Cetty \& Veron (2001) to calculate absolute B land
luminosity $L^{B}$. The value of $M^{nuc}_{B}$ is listed in Column
(3) in Table 1. The absolute B band magnitude ($M_{B}$) from
Veron-Cetty \& Veron (2001) is also listed in Column (2) in Table
1.

\section{Results}

We found in Table 1 that for PG quasars $M^{nuc}_{B}$ is almost
equal to $M_{B}$ , while for Seyfert 1 galaxies $M^{nuc}_{B}$ is
almost larger than $M_{B}$, namely, there is much contamination to
the luminosity of galactic nucleus from the host galaxies in
Seyfert 1 galaxies.

From Table 2, we suggested the reasonable value of $\alpha$ for
Seyfert 1 galaxies is about 0.1 $\sim$ 0.01 if the mean
inclination of Seyfert 1 galaxies is about 30 (deg) (Nandra et al.
1997; Wu \& Han 2001). Because the uncertainties of $Q$ and the
BLRs size don't effect our results too much compared with that of
$\alpha$, we just determined our results uncertainties from the
uncertainty of $\alpha$. We used $\alpha=0.1$ and $\alpha=0.01$ to
calculate the accretion disk parameters for Seyfert 1 galaxies.
The mean values of these three parameters are listed in Table 1.
The mean inclination of Seyfert 1 galaxies is $30.4\pm4.5$ (deg),
which is consistent with AGNs unification schemes (Urry \&
Padovani 1995). However, for PG quasars, the reasonable value of
$\alpha$ is 1 considering the small mean inclination using smaller
values of $\alpha$. We fixed $\alpha=1$ and adopted the same
method in Bian \& Zhao (2002) to calculate these three parameters,
which are consistent with the results in Bian \& Zhao (2002).
Their uncertainties are from the uncertainties of the values of
$Q$ and BLRs sizes. The values of black hole mass, accretion
rates, accretion rates in units of the Eddington accretion rate,
and inclinations are listed in Column (5)-(8) in Table 1,
respectively.

From Table 1 and Table 2, we found that for most Seyfert 1
galaxies accretion rates are less than one sun mass per year and
the accretion rates in units of the Eddington accretion rates are
less than one. The higher accretion rates for Seyfert 1 galaxies
showed in Bian \& Zhao (2002) is due to the overestimated absolute
B band magnitude, which can explain the question on
super-Eddington rates found by Collin \& Hure (2001). Collin \&
Hure (2001) also suggested that half of the objects in the sample
of Kaspi et al. (2000) are accreting close to the Eddington rate
or at super-Eddington rates unless BLRs is a flat thin rotating
structure with the same axis as the accretion disk, close to the
line of sight. Here we showed that the almost face-on BLRs,
namely, small inclination, indeed lead to the sub-Eddington rates
in the sub-sample of PG quasars.

In equation (2), we assumed that the gravitational instability of
the accretion disk leads to BLRs, and that the BLRs size
($R_{BLR}$) is equal to the value of the gravitational instability
radius ($R_{ins}$). It is possible that $ R_{ins} \ll R_{BLR}$.
Here we also calculated equation (1-3) assuming
$R_{ins}=0.1R_{BLR}$. Results are listed in Table 3. The
inclinations are consistent with that when we assumed
$R_{ins}=R_{BLR}$, while the accretion rates are larger. Equation
(2) is not directly related to the inclination. The relation
between $R_{ins}$ and $R_{BLR}$ would affect mainly the accretion
rates, not the inclinations. We should notice that we used the
BLRs sizes based on H$\beta$ emission line from the reverberation
mapping method. Peterson \& Wandel (2000) found variance emission
lines spanning an order of magnitude in distance from the central
source follows the expected $V \propto r^{-1/2}$ relation between
the emission line size and the emission line width, and they also
suggested that the gravity control broad emission line region
clouds, not the radiation pressure. Based on our calculation, we
found the gravitational instability would lead to the BLRs clouds
and $R_{ins}\approx R_{BLR}$.

In equation (3), we omitted the random isotropic velocity and
adopted the assumption of randomly-orientated BLR orbits (Bian \&
Zhao 2002). McLure \& Dunlop (2001; 2002) modelled the H$\beta$
FWHM distribution and suggested the disk-like BLRs model is
suitable for AGNs with FWHM larger than 2800 $km s^{-1}$. At the
same time, the smaller observed H$\beta$ line width in AGNs is
possibly from the isotropic velocity (Zhang \& Wu 2002). The disk
and isotropic figuration of the BLRs should be considered in this
case. We should be cautious of our results for AGNs with small
FWHM.

\section{conclusion}
Using the nuclear absolute B band magnitude instead of the total
absolute B band magnitude, we recalculated the central black hole
masses, the accretion rates, and the disk inclinations to the line
of sight according the method of Bian \& Zhao (2002). The main
conclusions can be summarized as follows:
\begin{itemize}
\item{Smaller nuclear absolute B band luminosity in Seyfert 1
galaxies compared with that from Veron-cetty \& Veron (2001) make
us to obtain different results from that of Bian \& Zhao (2002).}
\item{The value of $\alpha$ for Seyfert 1 galaxies and quasars is
different if the mean inclination of Seyfert 1 galaxies is about
30 (deg). The value of $\alpha$ for Seyfert 1 galaxies is about
0.1 $\sim$ 0.01, while the value of $\alpha$ for PG Quasars is
about 1. The difference of $\alpha$ maybe lead to the different
properties between Seyfert 1 galaxies and Quasars. }
\item{Choosing smaller value of $\alpha$ compared with that in
Bian \& Zhao (2002), we found that for most Seyfert 1 galaxies are
not accreting at super-Eddington rates. The higher accretion rates
for Seyfert 1 galaxies showed in Bian \& Zhao (2002) is due to the
adopted overestimated absolute B band magnitude.}

\end{itemize}

\begin{table*}
\begin{center}
\caption{The properties of the 37 AGNs. Col.1: name, Col.2:
absolute B band magnitude from Veron-Cetty et al. (2001), Col.3:
nuclear absolute B band magnitude, Col.4:log of the reverberation
mapping BH mass in $\Msolar$ from Kaspi et al. (2000), Col.5:log
of our calculated BH mass in $\Msolar$ , Col.6: log of accretion
rates in $\Msolar/yr$ , Col.7 the accretion rate in units of
Eddington accretion rate, Col.8: calculated inclinations (in deg)
to our sight.The nuclear absolute B band magnitude of Seyfert 1
galaxies labelled with $^{a}$ are from Ho $\&$ Peng (2001), the
rest of Seyfert 1 galaxies are from Ho (2002). The nuclear
absolute B band magnitude of Quasars are from Schmidt \& Green
1983.}
\begin{tabular}{lccccccccc}
\hline \hline

 Name      &$M_{B}$ & $M^{nuc}_{B}$&  $log_{10}M_{rm}$ &  $log_{10}M_{cal}$  &  $log_{10}\dot{M}$  & $\dot{m}$  & $i$  \\
           &(Mag)   & (Mag)        &  $(\Msolar)$      & $(\Msolar)$         &  $(\Msolar/yr)$     &            &(deg) \\
(1)&(2)&(3)&(4)&(5)&(6)&(7)&(8)\\
 \hline
3C120     &   -20.8 &  -20.79 &  7.36 &  9.00   $\pm$0.38 &  -1.22 $\pm$0.38 & 0.0045  $\pm$0.0043&5.4  $\pm$2.2   \\
3C390.3   &   -21.6 &  -21.22 &  8.53 &  8.88   $\pm$0.35 &  -0.76 $\pm$0.40 & 0.017   $\pm$0.016 &24.9 $\pm$10.0   \\
Akn120    &   -22.2 &  -22.62 &  8.26 &  9.60   $\pm$0.38 &  -0.71 $\pm$0.38 & 0.0037  $\pm$0.0035&7.8  $\pm$3.2   \\
F9        &   -23.0 &  -23.13 &  7.90 &  9.34   $\pm$0.38 &  -0.15 $\pm$0.38 & 0.025   $\pm$0.023 &6.94 $\pm$2.8   \\
IC4329A   &   -20.1 &  -19.25 &  6.70 &  6.76   $\pm$0.30 &   0.28 $\pm$0.43 & 23.63     $\pm$22.07   &36.9 $\pm$14.2  \\
Mrk79     &   -20.9 &  -19.93 &  7.72 &  8.27   $\pm$0.36 &  -0.95 $\pm$0.39 & 0.045   $\pm$0.043 &19.6 $\pm$7.9   \\
Mrk110    &   -20.6 &  -19.40 &  6.75 &  8.09   $\pm$0.38 &  -1.13 $\pm$0.38 & 0.046   $\pm$0.044 &7.8  $\pm$3.2   \\
Mrk279$^{a}$    &   -21.2 &  -20.55 &  7.62 &  8.18   $\pm$0.36 &  -0.50 $\pm$0.39 & 0.16   $\pm$0.15  &19.4  $\pm$7.9   \\
Mrk335$^{a}$    &   -21.7 &  -18.18 &  6.80 &  7.59   $\pm$0.37 &  -1.35 $\pm$0.39 & 0.09   $\pm$0.08 &14.8   $\pm$6.0   \\
Mrk509    &   -23.3 &  -22.48 &  7.76 &  9.92   $\pm$0.38 &  -1.12 $\pm$0.38 & 6.8E-4  $\pm$6.4E-4&3.00 $\pm$1.2   \\
Mrk590$^{a}$    &   -21.6 &  -16.46 &  7.25 &  7.20   $\pm$0.28 &  -1.78 $\pm$0.45 & 0.07   $\pm$0.067 &41.7  $\pm$15.5  \\
Mrk817$^{a}$    &   -22.3 &  -17.81 &  7.64 &  7.56   $\pm$0.27 &  -1.29 $\pm$0.45 & 0.1    $\pm$0.09 &44.3   $\pm$16.2  \\
NGC3227$^{a}$   &   -18.7 &  -16.01 &  7.59 &  7.17   $\pm$0.06 &  -1.27 $\pm$0.60 & 0.22    $\pm$0.20  &72.4 $\pm$12.2  \\
NGC3516$^{a}$   &   -20.5 &  -17.21 &  7.36 &  7.09   $\pm$0.17 &  -0.93 $\pm$0.52 & 0.63   $\pm$0.58  &57.7  $\pm$17.2  \\
NGC3783   &   -19.7 &  -19.01 &  6.97 &  7.24   $\pm$0.34 &  -0.43 $\pm$0.41 & 1.56     $\pm$1.46   &27.8   $\pm$11.1  \\
NGC4051$^{a}$   &   -16.8 &  -14.97 &  6.11 &  6.07   $\pm$0.28 &  -1.56 $\pm$0.44 & 1.67     $\pm$1.55   &41.2   $\pm$15.4  \\
NGC4151$^{a}$   &   -18.7 &  -19.18 &  7.18 &  7.13   $\pm$0.31 &  -0.15 $\pm$0.43 & 3.79     $\pm$3.54   &36.4   $\pm$14.0  \\
NGC4593   &   -19.7 &  -17.80 &  6.91 &  6.84   $\pm$0.27 &  -0.60 $\pm$0.45 & 2.55    $\pm$2.37   &42.8    $\pm$15.8  \\
NGC5548$^{a}$   &   -20.7 &  -17.29 &  8.09 &  7.77   $\pm$0.14 &  -1.45 $\pm$0.54 & 0.039   $\pm$0.036 &61.9 $\pm$16.4  \\
NGC7469$^{a}$   &   -21.6 &  -17.78 &  6.81 &  6.89   $\pm$0.31 &  -0.75 $\pm$0.42 & 1.669     $\pm$1.559   &35.3 $\pm$13.7  \\
PG0026    &   -24.0 &  -24.35 &  7.73 &  9.645  $\pm$0.74 &  0.277 $\pm$0.74 & 0.17027  $\pm$0.16991   &4.24    $\pm$0.64    \\
PG0052    &   -24.5 &  -23.99 &  8.34 &  9.631  $\pm$0.75 &  0.082 $\pm$0.76 & 0.11803  $\pm$0.11781   &9.12    $\pm$1.40    \\
PG0804    &   -23.9 &  -23.17 &  8.28 &  9.415  $\pm$0.70 &  -0.194$\pm$0.70 & 0.07901  $\pm$0.07876   &9.39    $\pm$1.33    \\
PG0844    &   -23.1 &  -23.30 &  7.33 &  8.443  $\pm$0.86 &  0.857 $\pm$0.87 & 18.06383 $\pm$18.05147  &10.09   $\pm$1.77    \\
PG0953    &   -25.6 &  -25.24 &  8.26 &  10.114 $\pm$0.74 &  0.341 $\pm$0.74 & 0.06530  $\pm$0.06515   &3.91    $\pm$0.59    \\
PG1211    &   -24.0 &  -23.31 &  7.61 &  9.195  $\pm$0.79 &  0.104 $\pm$0.80 & 0.40511  $\pm$0.40458   &5.77    $\pm$0.93    \\
PG1226    &   -26.9 &  -26.47 &  8.74 &  11.066 $\pm$0.72 &  0.126 $\pm$0.72 & 0.00416  $\pm$0.00415   &2.51    $\pm$0.37    \\
PG1229    &   -22.4 &  -22.61 &  7.88 &  8.564  $\pm$0.91 &  0.338 $\pm$0.92 & 5.21516  $\pm$5.21290   &15.66   $\pm$2.88    \\
PG1307    &   -24.6 &  -24.07 &  8.45 &  9.422  $\pm$1.00 &  0.340 $\pm$1.01 & 1.09526  $\pm$1.09505   &9.82    $\pm$1.99    \\
PG1351    &   -24.1 &  -22.52 &  7.66 &  9.408  $\pm$0.88 &  -0.584$\pm$0.89 & 0.07716  $\pm$0.07711   &4.62    $\pm$0.83    \\
PG1411    &   -24.7 &  -22.95 &  7.90 &  9.075  $\pm$0.85 &  0.013 $\pm$0.86 & 0.56902  $\pm$0.56858   &8.97    $\pm$1.55    \\
PG1426    &   -23.4 &  -22.91 &  8.67 &  9.023  $\pm$0.85 &  0.091 $\pm$0.87 & 0.80285  $\pm$0.80228   &23.31   $\pm$4.00    \\
PG1613    &   -23.5 &  -23.43 &  8.38 &  8.794  $\pm$0.87 &  0.630 $\pm$0.89 & 4.98799  $\pm$4.98486   &22.96   $\pm$4.00    \\
PG1617    &   -23.4 &  -23.06 &  8.44 &  9.030  $\pm$0.79 &  0.150 $\pm$0.80 & 0.66211  $\pm$0.66124   &18.12   $\pm$2.89    \\
PG1700    &   -25.8 &  -25.46 &  7.78 &  9.045  $\pm$2.14 &  1.542 $\pm$2.14 & 10.52923 $\pm$10.52919  &3.67    $\pm$1.49    \\
PG1704    &   -25.6 &  -25.50 &  7.57 &  10.146 $\pm$1.36 &  0.464 $\pm$1.36 & 1.41944  $\pm$1.41943   &1.13    $\pm$0.32    \\
PG2130    &   -22.9 &  -22.61 &  8.16 &  9.396  $\pm$0.75 &  -0.512$\pm$0.75 & 0.05085  $\pm$0.05075   &9.21    $\pm$1.40    \\
\hline
\end{tabular}
\end{center}
\end{table*}

\begin{table*}
\begin{center}
\caption{The mean and the standard deviation of inclinations in
degree, log of accretion rates in $\Msolar/yr$, and the accretion
rate in units of Eddington accretion rate for Seyfert 1 galaxies
and for PG quasars just considering $Q=1$ and different $\alpha$.}
\begin{tabular}{lccccccc}
\hline\hline
$\alpha$ &1&0.1&0.05&0.01&0.001\\
\hline
$i$ (Seyfert 1) & 62.8$\pm$27.3&40.7$\pm$25.1&35.9$\pm$23.9&20.1$\pm$15.4&8.7$\pm$7.0\\
$i$ (PG quasar) & 9.3$\pm$6.7&3.9$\pm$2.8&3.2$\pm$2.3&1.6$\pm$1.2&0.7$\pm$0.5\\
$log_{10}\dot{M}$(Seyfert 1)&0.6$\pm$0.5&-0.5$\pm$0.5&-0.3$\pm$0.9&-1.3$\pm$0.5&-2.1$\pm$0.6 \\
$log_{10}\dot{M}$(PG quasars)&0.1$\pm$0.4&-0.6$\pm$0.4&-0.5$\pm$0.8&-1.4$\pm$0.4&-2.1$\pm$0.4\\
$\dot{m}$ (Seyfert 1)&76.7$\pm$221&3.5$\pm$10.1&3.1$\pm$5.6&0.1$\pm$0.4&0.004$\pm$0.01\\
$\dot{m}$ (PG quasars)&0.06$\pm$0.1&0.002$\pm$0.004&0.5$\pm$2.0&5.6E-5$\pm$1.3E-4&1.7E-6$\pm$4.1E-6\\
\hline
\end{tabular}
\end{center}
\end{table*}

\begin{table*}
\begin{center}
\caption{The same as Table 2, but for $R_{ins}=0.1R_{BLR}$}
\begin{tabular}{lccccccc}
\hline\hline
$\alpha$ &1&0.1&0.05&0.01&0.001\\
\hline
$i$ (Seyfert 1) & 67.6$\pm$26.0&47.6$\pm$26.8&40.5$\pm$25.0&20.1$\pm$15.4&11.1$\pm$8.9\\
$i$ (PG quasar) & 11.9$\pm$8.5&5.0$\pm$3.6&3.9$\pm$2.8&1.6$\pm$2.1&0.9$\pm$0.6\\
$log_{10}\dot{M}$(Seyfert 1)&1.9$\pm$0.6&0.8$\pm$0.5&0.5$\pm$0.5&-0.1$\pm$0.5&-0.9$\pm$0.6 \\
$log_{10}\dot{M}$(PG quasars)&1.4$\pm$0.35&0.6$\pm$0.4&0.4$\pm$0.4&-0.2$\pm$0.4&-0.9$\pm$0.4\\
$\dot{m}$ (Seyfert 1)&1.7E+4$\pm$5.1E+4&876$\pm$2537&333$\pm$962&32.2$\pm$92.3&1.1$\pm$2.9\\
$\dot{m}$ (PG quasars)&16.0$\pm$37.8&0.5$\pm$1.2&0.4$\pm$0.4&-0.2$\pm$0.4&4.66E-4$\pm$1.1E-3\\
\hline
\end{tabular}
\end{center}
\end{table*}

\begin{acknowledgements}
We thank the anonymous referee for the valuable comments. This
work has been supported by the NSFC (No. 10273007) and NSF from
Jiangsu Provincial Education Department.
\end{acknowledgements}

\end{document}